\documentclass{aa}
\begin{document}

   \thesaurus{06     
              (03.11.1;  
               16.06.1;  
               19.06.1;  
               19.37.1;  
               19.53.1;  
               19.63.1)} 
   \title{WSRT Observations of the Hubble Deep Field Region}


   \author{M.A. Garrett\inst{1}, A.G. de Bruyn\inst{2,4},
           M. Giroletti\inst{1,3}, W.A. Baan\inst{2} 
           \and R.T. Schilizzi\inst{1}}

   \offprints{M.A. Garrett}

   \institute{Joint Institute for VLBI in Europe (JIVE),  
              Postbus 2, 7990~AA, Dwingeloo, The Netherlands 
              (garrett@jive.nl).
   \and
   Netherlands Foundation for Research in Astronomy (ASTRON/NFRA), 
              Postbus 2, 7990~AA, Dwingeloo, The Netherlands.
  \and Dipartimento di Fisica dell Universita di Bologna,
   via Irnerio 46, I-40126, Bologna, Italy.   \and
   Kapteyn Astronomical Institute, Postbus 800, 9700 AV Groningen, 
              The Netherlands.}


   \authorrunning{Garrett et al.} 
   \titlerunning{WSRT 1.4 GHz Observations of the HDF} 

   \maketitle

   \begin{abstract}
     
     We present deep WSRT 1.4 GHz observations of the Hubble Deep Field
     region. At the $5\sigma$ level, the WSRT clearly detects 85
     regions of radio emission in a $10^\prime \times10^{\prime}$ field
     centred on the HDF. Eight of these regions fall within the HDF
     itself, four of these are sources that have not previously been
     detected at 1.4 GHz, although two of these are VLA detections at
     8.5~GHz. The two new radio sources detected by the WSRT are
     identified with relatively bright ($I<21^{m}$) moderate redshift
     spiral and irregular type galaxies.  In the full field, the WSRT
     detects 22 regions of emission that were not previously detected
     by the VLA at 1.4~GHz.  At least two of these are associated with
     nearby, extended star-forming galaxies.

   \end{abstract}

%

\section{Introduction}

Observations of the Hubble Deep Field (HDF) region (\cite{W96},
hereafter W96) at centimeter wavelengths are now advancing our
understanding of the nature of the faint, microJy radio source
population (\cite{R98}, hereafter R98; \cite{R99}, hereafter R99;
\cite{R00}, hereafter R00, and \cite{M99}, hereafter M99).  These
observations suggest that $\sim 70$\% of faint sub-mJy and microJy
radio sources are identified with low radio luminosity
($L<10^{23}$W/Hz), steep spectrum, moderate redshift ($z\sim 0.2-1$)
star forming galaxies. These sources are often identified with
morphologically peculiar, merging or interacting galaxies, with blue
colours, and HII-like emission spectra (R98). The remaining 20\% of the
faint radio population are identified with relatively low-luminosity
AGN and 20\% are optically faint sources with no detections down to
I=25.5 in the HFF and I=28.5 in the HDF. These
optically faint systems are thought to be distant galaxies, obscured by
dust (R99).  The vast majority of far infrared ISO detections in the
HDF (and the adjacent Hubble Flanking Fields, HFF) are also detected in
the radio (\cite{A99}, hereafter A99), suggesting that the same strong
correlation between the far infrared and radio continuum flux densities
(as observed for nearby star-forming galaxies - see \cite{C92} and
references therein) also holds for these fainter, more distant systems 
\cite{B00}.

In order to further advance the study of the faint microJy radio source
population, we observed the HDF and the surrounding flanking fields
(HFF) with the recently upgraded Westerbork Synthesis Radio Telescope
(WSRT) at 1.4 GHz.

\section{WSRT Observations and Data Analysis}

Observations of the HDF region were made with the upgraded WSRT at 1.38
GHz in the period from 20 April until 6 May 1999.  A total of six
(uninterrupted) 12 hour observations were made with different array
configurations resulting in an excellent uv-coverage with baselines
ranging from 36 to 2760 meters (with an increment of 12 meters).  The
WSRT continuum back-end provided 8 contiguous 10 MHz bands running from
1340-1420 MHz. The visibilities were averaged over 60 seconds.  Full
polarization information was obtained but is not presented here.

\begin{figure*}
\label{wsrt_hdf_hff} 
\vspace{11cm}
\begin{picture}(150,190)
\put(120,-50){\includegraphics{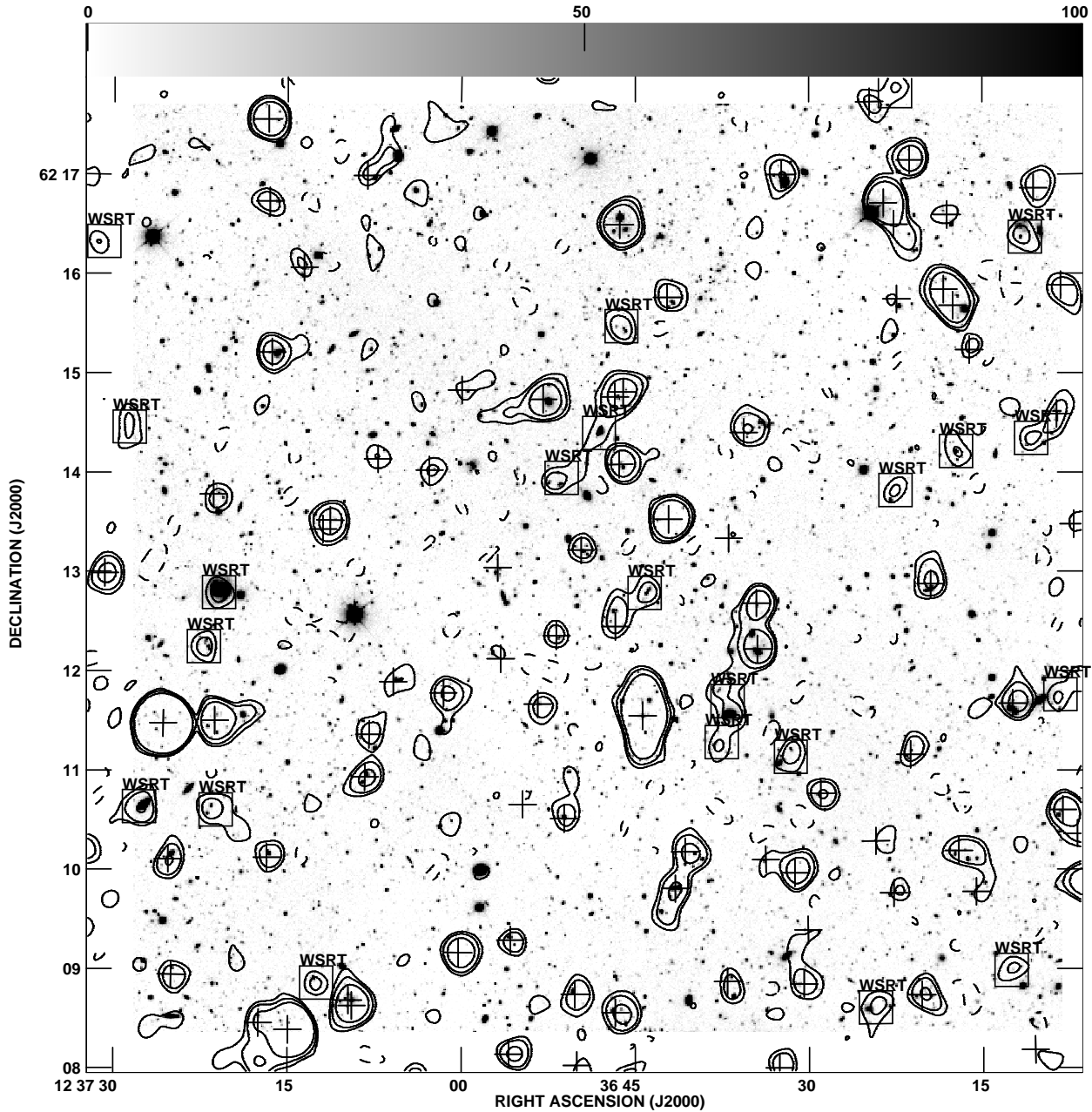}}
\end{picture}
\caption{The WSRT image of the HDF (inner $2.5^{\prime}$) and HFF regions 
  (a $10^{\prime\prime}\times10^{\prime\prime}$ field), restored with a 
  15$^{\prime\prime}$ circular beam and superimposed upon the deep CFHT
  I-band image of \cite{B98}.  The $1\sigma$ rms noise level is $\sim 9$
  microJy/beam with contours drawn at -3, 3, 5 and $10\sigma$.  Crosses
  represent sources detected by the VLA at similar rms noise levels
  (R00).  Sources detected by the WSRT (but not the VLA or MERLIN at
  1.4~GHz) are boxed and appropriately labeled.}
\end{figure*}

The data analysis was performed in NEWSTAR (\cite{N93}) following the
standard WSRT processing route. 3C286 was observed once per 12 hour
run, and the complex gain solutions thus determined, were transferred
to the HDF data. Typically, about 90\% of the data proved to be of good
and usable quality.  The standard WSRT taper was applied to the data
and baselines shorter than 100 meter were not used in the image,
effectively filtering out large scale $> 5^{\prime}$ emission features
of both instrumental and galactic (foreground) origin. Images of $2048
\times 2048$ pixels covering an area of $2.8^{\circ}\times2.8^{\circ}$
were generated. The brightest 121 sources in the field, with flux
densities ranging from a high of 130 mJy down to 0.35 mJy, were used to
self-calibrate the data (with a 60 second solution interval), and then
subtracted from the uv-data.  The central $900 \times 900$ pixels in
the residual image was then cleaned down to a level of 28 $\mu$Jy
yielding 8000 clean components. These were restored with a Gaussian
beam with half-width $14.3^{\prime\prime} \times 15.5^{\prime\prime}$
(RA $\times$ DEC). On top of this image we restored the
self-calibration model, after first multiplying the residual image with
a factor 1.2 to correct for self-calibration noise bias (\cite{W92}).
The full field has a r.m.s. noise level that varies from about 7.5 to
$8\mu$Jy/beam, almost identical to the $7.5\mu$Jy noise level of the
1.4 GHz VLA image (R00). The local noise level around the strongest
off-axis sources is affected by (we presume) pointing problems,
however, all these sources are located outside the central
$10^{\prime}\times10^{\prime}$ region. Towards the centre of the image
the noise appears to go up slightly. We attribute this to source
confusion estimated at $5\mu$Jy/beam which, when added in quadrature,
raises the noise level to about $9\mu$Jy/beam in the central region of
the field (this noise component is not significant beyond a radius of
$0.3^{\circ}$ due to the primary beam attenuation).  Support for this
interpretation comes from the extremely uniform noise level of
$7.5\mu$Jy in the Stokes Q image.  

The final total
intensity image of the central HDF and HFF region is presented in
Fig.\ref{wsrt_hdf_hff}, superimposed upon the deep CFHT I-band image of
\cite{B98}. This $10^{\prime}\times10^{\prime}$ WSRT image is corrected
for primary-beam attenuation and is centred on the HDF. The HDF/HFF
WSRT image was exported from NEWSTAR to AIPS and source positions were
fitted using the task {\sc IMFIT}. Regions of emission with a measured
peak flux density in excess of the assumed $5\sigma$ noise level
($45\mu$Jy) are considered to be bona fide detections. Note that the
most negative feature in the analysed area has a flux density of
$-42\mu$Jy. In principle, the accuracy with which we can determine the
absolute position of the faintest unresolved radio sources in the field
is $\sim 1.5^{\prime\prime}$ (1/10 of the restoring beam).  For
extended regions of emission this formal error is rather optimistic,
and for the purposes of optical identifications, we adopt a positional
error of $3^{\prime\prime}$.

\section{Radio Sources in the central HDF region} 

Within the central HDF region the WSRT detects eight radio sources.
Their positions, flux densities and sizes are presented in table 1. In
most cases the deconvolved sizes (major axis, minor axis, PA) are
upper-limits since the sources are smaller than the WSRT beam. The 1.4
and 8.5~GHz VLA flux densities (R98, R00), ISO ($15\mu$m) flux
densities (A99), and the spectroscopic redshifts (\cite{C00}, hereafter
C00) of the sources are also presented.

\begin{table*} 
\caption{WSRT 1.4 GHz HDF Source List}
\begin{tabular}{|l|l|l|l|l|l|l|l|l|l|} 
\label{table1} 
Source & RA ($+ 12$ hr) & DEC ($+ 62^{\circ}$) &   S$_{P}$ & S$_{T}$ &
Size & S$_{VLA_{1.4}}$ & S$_{VLA_{8.5}}$ & S$_{15\mu}$ & $z$ \\ 
       & (h,m,s)        & ($^{\circ}$ $^{\prime}$ $^{\prime\prime}$) 
& $\mu$Jy & $\mu$Jy & ($^{\prime\prime}$ $\times$ $^{\prime\prime}$, 
$^{\circ}$) & $\mu$Jy &$\mu$Jy & $\mu$Jy & \\ 
\hline 

3644+1247 &  36 44.190 &  12 47.31  & $67\pm7  $ & $73\pm12$ & $<13$ & 
- & 10 & 282 & 0.555 \\

3644+1133 &  36 44.229 &  11 33.68 & $1190\pm14$ & $ 1606\pm14$ &  
$< 15$ & 1290 & 477 & - & 1.050 \\ 

3646+1405 &  36 46.019 &  14 05.63 & $187\pm5$ & $ 187\pm5$  & $< 15$ 
& 179 & 168 & 107 & 0.962 \\

3646+1236 &  36 46.284 & 12 36.03 & $64\pm6$  & $167\pm22$  &
$32\times6$, 140 & - & - & - & 0.321  \\ 

3647+1427 & 36 47.839 & 14 27.02 & $47\pm6$ & $116\pm21$ & $< 50$  
& - &  9.8 & 307 & 0.139 \\ 

3649+1314 & 36 49.575 & 13 14.12 & $68\pm4$ & $68\pm4$ & $ < 15$  
& 49 & 14 & 320 & 0.475 \\ 

3651+1357 & 36 51.359 & 13 57.04 & $64\pm7$ & $82\pm14$ & $< 18$  
& - & - & 151 & 0.557 \\ 

3651+1222 & 36 51.811 & 12 22.37 & $57\pm5$ & $57\pm5$ & $< 15$  
& 49&  16 & 48 & 0.401 \\ 
\hline
\end{tabular} 
\end{table*} 

Five of these detections (3644+1247, 3644+1133, 3646+1405, 3649+1314,
and 3651+1222), are clearly associated with sources detected at the
$5\sigma$ level by either the VLA at 8.5~GHz (R98) or the VLA/MERLIN at
1.4~GHz (see R00, M99 \& Table 1). Their optical identifications are
discussed in R98. In addition, 3647+1427 is clearly associated with the
$4\sigma$ VLA 8.5~GHz source 3648+1427 (from the supplemental radio
source catalogue of R98). Note that there is no formal VLA 1.4~GHz
detection of 3644+1247, 3646+1236, 3647+1427, or 3651+1357.  In the
remainder of this paper we attempt to determine optical identifications
for sources in the HDF region that are only detected by the WSRT. We
employ the standard likelihood ratio (LR) analysis of \cite{M2000} (and
references therein).  We have compared the WSRT radio source positions
in the HDF with the HDF HST Catalogue (W96), and in the HFF with the
deep I-band catalogue of \cite{B98} (hereafter B98). The spectroscopic
redshifts referred to throughout this paper are taken from \cite{C00}.
We assume an error in the source positions of 3$^{\prime\prime}$ in the
radio and 0.5$^{\prime\prime}$ in the optical.

One very extended region of emission, 3646+1236, is not detected by the
VLA at 8.5~GHz but some fraction of the WSRT emission may be 
associated with VLA J123646+621226. However, the centroid of this
extended region lies 10$^{\prime\prime}$ to the north of the 1.4 GHz
VLA source. The most likely optical identification of 3646+1236
(identification probability, $P> 72$\%) is with 4-241 (W96), a bright
($I \sim 20.6^{m}$), irregular (possibly merging), $z=0.321$ galaxy
lying within $\sim 3.5^{\prime\prime}$ of the radio centroid. Although
there is no ISO detection in this area, the morphology of the optical
identification suggests that the steep spectrum radio emission ($\alpha
< -1$) is generated by star formation.  If half the total radio
emission in this area is associated with 3646+1236 itself, then
following \cite{H1999} (and assuming a Salpeter IMF), we derive an
upper-limit to the K-corrected star formation rate (SFR) of
17~M$_{\odot}$yr$^{-1}$ (taking $q_{0}=0.5$, H$_{0}= 50$ km/sec/Mpc).

3651+1357 is not detected by either the VLA at 8.5~GHz or the
VLA/MERLIN at 1.4~GHz.  The most likely optical identification ($34\%<
P <60$\%) is 2-652.0 (W96), a bright ($I\sim 20.6^{m}$), Sbc galaxy
lying 4.4$^{\prime\prime}$ to the south-east of the radio centroid with
a spectroscopic redshift $z=0.557$. There is an ISO $15\mu$m detection
in this area (A99) also identified with 2-652.0 (\cite{M97}). The
non-detection of 3651+1357 in the 10$^{\prime\prime}$ resolution
8.5~GHz VLA image (R98), implies a steep radio spectral index ($\alpha
\sim -1$). The coincident WSRT and ISO detections suggest that 2-652.0
is a star-forming galaxy. We derive an upper-limit to the K-corrected
SFR of $38$~M$_{\odot}$yr$^{-1}$, in good agreement with
the ISO estimate (\cite{RR97}).

There is one source in the HDF central region (VLA J123656+621302, a
$z=0.474$ elliptical galaxy, R98) that is detected by the VLA at 1.4
and 8.5~GHz ($S_{1.4}\sim49.5\mu$Jy, S$_{8.5}=8\mu$Jy) but not by the
WSRT, even at the $3\sigma$ level (the WSRT peak flux in this region is
$13\mu$Jy/beam). An inspection of each of the six 12 hour WSRT runs
shows no emission in this region, suggesting that this source may be
variable on a time-scale of months or years. R98 identify this source
as a probable low-luminosity AGN. Very little is known about the radio
variability properties of faint AGN, but assuming such sources are
self-absorbed, for a given magnetic field strength, fainter sources
should also be smaller sources. In the $10^{\prime}\times10^{\prime} $
field considered here, only one other VLA 1.4 GHz detection (VLA
J123654+621039, S$_{1.4}=48\mu$Jy) is not detected by the WSRT below
the $3\sigma$ level (the WSRT peak flux density in this region is
$22\mu$Jy). We have been unable to identify a plausible optical
identification for this source from the VLA position.  Given that radio
variability in distant star-forming galaxies seems unlikely, this
source is also a potential AGN candidate.

\subsection{Detections in the HFF Region} 

A complete radio catalogue of sources in the WSRT HDF and HFF region
(including optical identifications) is in preparation (a preliminary
catalogue is online at \verb+www.jive.nl/~mag/hdf+).  85 distinct
regions of emission are clearly detected with the WSRT above the
$5\sigma$ limit within a $10^{\prime}\times10^{\prime}$ field centred
on the HDF. Of these 85 regions of emission, 55 are associated with
discrete, single component VLA 1.4 GHz sources, 8 are associated with
multiple VLA 1.4~GHz sources (for the WSRT these blend together to form
a complex region of extended emission), and 22 are clearly detected by
WSRT alone. As discussed earlier, some fraction of these WSRT-only
detections are blends of sources that are presumably resolved by the
higher resolution VLA/MERLIN 1.4 GHz observations.  However, some 
fraction of the WSRT-only detections are clearly discrete
sources and some of these are identified with bright, nearby galaxies.

An inspection of Fig~\ref{wsrt_hdf_hff} highlights two clear cases of
this: 3720+1247 - coincident and clearly identified ($P>99$\%) with an
$I\sim 18.4^{m}$, z=0.106 spiral galaxy, and 3636+1132 - coincident and
clearly identified ($P>99$\%) with a $I\sim 18.6^{m}$, z=0.078 spiral
galaxy. The latter source falls within the region surveyed by ISO and
is coincident with a ISO $15\mu$m detection (A99). These two sources
have radio powers that are a factor of $\sim 2$ higher than the
typically brightest extragalactic radio supernovae (Wilkinson \& de
Bruyn, 1990) and in principle, they could be recent radio SNe that have
sharply increased in flux density over the 2.5 years that separate the
VLA and WSRT observations. However, we consider it more likely that
these detections are cases of resolved extended disk emission. If all
the radio emission associated with these sources is due to star
formation, we estimate modest SFRs of $5$~M$_{\odot}$yr$^{-1}$ for
3720+1247 and $3$~M$_{\odot}$yr$^{-1}$ for 3636+1132. The fact that
they are not detected by the VLA at 1.4~GHz, suggests that the radio
emission could arise from a substantial area of the optical disk ($ >
5-10$~kpc across).

\section{Conclusions}

Deep 1.4~GHz WSRT observations have detected 2 new radio sources in the
HDF. These are associated with moderate redshift star-forming galaxies.
In the full $10^{\prime}\times10^{\prime}$ field, 22 new regions of
radio emission are detected by the WSRT that are presumably resolved
out by the VLA/MERLIN. At least two of these are clearly associated with
relatively bright, nearby star-forming galaxies with modest SFRs.
Further analysis of the full field is required to distinguish how many
of the remaining regions are associated with discrete, extended radio
sources, and how many are blends of closely separated, faint radio sources
confused in the WSRT beam.  The non-detection of 2 sources previously
detected by the VLA at 1.4~GHz, hints at variability in the microJy AGN
radio source population.

\begin{acknowledgements}
The WSRT is operated by the Netherlands Foundation for Research in
Astronomy (ASTRON) with financial support by the Netherlands
Organisation for Scientific Research (NWO).
\end{acknowledgements}

\end{document}